# Fan-shaped and toric textures of mesomorphic oxadiazoles

A. Sparavigna[1], A. Mello[1], and B. Montrucchio[2]

[1] Dipartimento di Fisica, Politecnico di Torino

[2] Dipartimento di Automatica ed Informatica, Politecnico di Torino

C.so Duca degli Abruzzi 24, Torino, Italy

**Abstract**

When a family of non symmetrical heterocycled compounds is investigated, a variety of mesophases can be observed with rather different features. Here we report the behaviour of seven different members among a family of such materials, that consists of mesomorphic oxadiazole compounds. In two of these compounds, the optical microscope investigation shows very interesting behaviours. In their smectic phases, fan-shaped and toric textures, sometimes with periodic instability, are observed. Moreover, the nematic phase displays a texture transition.

Texture transitions have been previously observed only inside the nematic phase of some compounds belonging to the families of the oxybenzoic and cyclohexane acids. In these two oxadiazole compounds we can observe what we define as a "toric nematic phase", heating the samples from the smectic phase. The toric nematic texture disappears as the sample is further heated, changing into a smooth texture.

**Introduction**

The correlation between the existence of mesophases and their features and the chemical structure of molecules is the core of the research in liquid crystals. As we have seen studying some alkyloxybenzoic acids, an increased length of the alkyl chain produces a strong increase of the nematic mesophase range [1,2]. These acids are materials where the mesogenic unit structure is a symmetrical hydrogen-bonded dimer. A simple binary mixture of two members of the same group of acids is able to enlarge the smectic range [3]. When a family of non symmetrical structures is under investigation, the variety of results is strongly enhanced. Here we discuss the mesomorphic behaviour of some members among a family of non symmetrical compounds, consisting of mesomorphic oxadiazole compounds [4-6]. These materials have been developed because their molecular structure with heterocycles as substituents gives smectic and nematic mesophases. It was observed in particular, that not only the chemical structure of the substituents, but also their position with respect to the oxadiazolic ring of the molecule is relevant for the mesophases.

The optical microscope investigations, which we are discussing in this paper, show very interesting behaviour of the smectic and the nematic phase of some of these oxadiazole compounds. In the smectic phases, fan-shaped and toric textures, with periodic instability in the toric structure, are observed. Moreover, it is also remarkable the behaviour of the nematic phase, displaying a texture transition. These transitions have been previously observed inside the nematic phase of some mesomorphic thermotropic materials belonging to the families of the alkyloxybenzoic and cyclohexane acids [7-13]. It is a more or less abrupt change, driven by the temperature, in the nematic texture seen by means of the polarised light microscope. In some cases, it is also possible to observe the transition in the calorimetric investigations and dielectric spectroscopy [8,14]. In other cases, it is necessary an image processing aided procedure to enhance the observation of the transition [1,9].

Alkyloxybenzoic and cyclohexane acids are mesogenic because the mesogenic units are dimers. The more common explanation for the presence of a texture transition in the nematic range is based on the existence of cybotactic clusters of dimers, favouring a local smectic order in the nematic range [15,16]. Our observation of a texture transition inside the nematic phase of two oxadiazole compounds suggests a possible existence of cybotactic clusters in these materials too. We start with a short description of the oxadiazole family used for investigations and then the more deep discussion of the smectic phases and of the texture transitions.

**The oxadiazole compounds.**

The members of the oxadiazole family we are now discussing are those whose structures are reported in Fig.1. These compounds have a non symmetrical heterocycled structure. They have been previously studied with the DSC differential scanning calorimetry [5]. As it is possible to see in Table 1, which reports the transition temperatures, the molecular structure is strongly influencing the mesophase ranges. Not only the presence of benzene or cyclohexane rings, but also their position, with respect to the thiadiazolic or imidazolic part of the molecule, is important for the existence of mesophases.

All the samples were inserted in the cell when the material was in the isotropic phase. The walls of liquid crystal cells are untreated clean glass surfaces. No treatments were done to favor planar or homeotropic alignments. The liquid crystal cells are heated and cooled in a thermostage and textures observed with a polarized light microscope. Let us start the discussion from compound A of Figure 1: it has only the smectic phase.

**Compound A**

It is an interesting material, that displays at room temperature the texture shown in the upper part of Figure 2. In Table I, we guess the phase at room temperature to be a crystal phase, or a rigid smectic phase. On heating, the sample changes into a smectic phase with a fan-shaped texture at the temperature of 140 °C. The material becomes an isotropic liquid at the temperature of 210°C. The material does not show a nematic phase. On cooling the sample has a more complex behaviour. First of all, beautiful bâtonnets appears cooling from the isotropic liquid, with periodic instabilities insides. Then larger domains grow, and the texture changes very fast in the image frame as in a scene played by moving fans. The smectic phase has a strong hysteresis on cooling, because the fan-shaped texture is observed till down 110°C. Then the sample probably becomes a crystal or a rigid smectic F. The transition is shown in Fig.3. A further reduction of the temperature gives the texture in Figure 2. In this rigid phase observed with crossed polarisers, the crystal shows thick dark lines among the lamellar structure. These lines are cracks as it is possible to see when they arise as the sample is cooled down from the smectic phase.

**Compound B**

On heating, we observed the following temperature transitions: 156°C from crystal to smectic, 172°C from smectic to nematic and 200°C from the nematic to the isotropic phase. On cooling, we have the transition from the isotropic liquid to the nematic phase at 198°C and the transition into the smectic phase at 168°C. The crystal phase appears at 145°C. This compound possesses a nematic phase too. The smectic phase has not only the fan-shaped texture but also regions with macroscopic toric arrangement of the smectic planes. The toric domains are focal conic domains where the ellipse is degenerated in a circle and the hyperbola is degenerated in a straight line [17].
If we look at the sequence in Figure 4, the nematic, the smectic and the crystal phases have textures where the same overall arrangement of defects is unchanged on cooling. It could be a memory effect of the surface layers of the material, close to the cell walls.

**Compound C**

The sample has a crystal phase till 94°C. Then a smectic phase till 172°C. We observed the clearing point at 233°C. On cooling, the nematic phase appears at 232°C, the smectic phase at 163°C, and the crystal at 65°C. What is remarkable in this compound is a texture transition inside the nematic phase. We can observe what we define as a "toric nematic phase" on heating till 195-200°C, and then a smooth nematic texture till the clearing point. On cooling, the toric phase appears around 200°C. This behaviour of the nematic phase is observed also in the compound D, that we will describe in the next section.

The smectic phase of sample C has a fan shaped and toric smectic phase. The Figure 5 shows in its upper part, the arrangement of the toric defects in the smectic phase. Sometimes on cooling the smectic phase, it happens that the toric texture displays inside an undulation instability (lower part of Fig.5). A recent theory of the lamellar phases could be invoked to explain the appearance of undulations in the planes of the lamellar phases [18,19].

Let us discuss more deeply the nematic phase. On heating, the nematic phase substitutes the smectic phase, the transition is shown in Fig.6. The nematic phase in the figure has a texture that shows a "toric" arrangement of the defects, and this is why we call this nematic phase a "toric nematic phase" . The texture remembers the shape of domains in the smectic phase. On heating, this nematic texture disappears between 195-200°C, depending on the position in the cell observed at microscope, and a smooth texture is displayed by the nematic. This is the texture that is observed on cooling from the isotropic phase. On cooling, around 200°C, the toric texture starts to grow in the cell as shown in the Figure 7.

It is remarkable that a compound, rather different from the dimeric acids [1,2], has a texture transition in the nematic phase. If we assume as a reasonable explanation for the existence of a texture transition, the presence in the nematic melt of cybotactic clusters with local smectic order, we can invoke the presence of such clusters also in this material. The local smectic order is responsible for the toric nematic texture.

**Compound D**

The sample has a crystal phase that changes in another crystal phase at 82°C. The crystal becomes to be smectic at 103°C. The smectic phase persists till 110°C. We observed the clearing point at 232°C. On cooling, the nematic phase appears at 230°C, the smectic phase at 97°C, and the crystal at 71°C. This compound, as the previous one, has a texture transition inside the nematic phase. We can observe the toric nematic phase on heating till 150°C, and then a nematic texture till the clearing point. On cooling, the toric phase appears around 150°C. The Figure 8 shows a beautiful toric nematic phase, with the bent structure of the defects. The bent contours between the nematic domains are clearly visible in this figure.

In fact, samples C and D are similar in the behaviour of the smectic and nematic phases. They have a very slight change in the structure of the molecule, as we can see in the Figure 1, with respect of the oxadiazolic group. The different position changes the transition temperatures.

**Compounds E,F and G**

These three compounds have only the nematic mesophase. In the nematic range there is no evidence of a texture transition. Sample E has a transition from a crystal phase in the nematic phase at 46°C, on heating. The clearing point is at 109°C. On cooling, the nematic phase appears at 107°C. The nematic phase remains till the room temperature and just after a very long relaxation time, the crystal is formed. The nematic phase has always a visible thermal flickering of the texture. The phase sequence for sample F is: a crystal phase till 72°C that changes into a nematic phase till 99°C. On cooling, the nematic starts from 98°C with a very fast transition. The sample remains nematic till 50°C. For sample G, we observed a transition from crystal to nematic at 80°C, and a short nematic range till 87°C. On cooling, the nematic range extends till 51°C. This behaviour, a wide nematic range on cooling, is common to these three materials and confirmed by previous DSC measurements.

**Discussion.**

As already observed in some members of the alkyloxybenzoic acid family (6-,7-,8-,9OBAC), in the binary mixtures of 6OBAC with other members of this family and in a cyclohexane acid (C6), the nematic phase exhibits a different order at low and at high temperatures. This different order is observed with the optical microscopy because the cell has two different textures in the nematic range. For the previously mentioned compounds, the low temperature nematic texture is composed of very small domains and looks like a granular structure. If the sample has a smectic phase, the texture of the smectic phase is also showing a granular texture.

In the case of the oxadiazole compounds C and D, we observe a texture transition in the nematic phase driven by the temperature, with a low temperature subphase textured with bent-shaped contours between domains. We defined this texture as a toric texture, because a further cooling of the sample gives a smectic phase with toric domains. This is consistent with the hypothesis that a concentration of cybotactic clusters is at the origin of the texture transition in the nematic phase. Let us remember that the cybotactic clusters are groups of molecules with a local smectic phase; they can be imagined as smectic seeds in the nematic fluid.

In a previous paper, we proposed a model to explain the origin of different textures in the nematic phase [1]. Let us imagine a thin liquid crystal cell in the smectic phase, with defects in the smectic structure. Rising the temperature, the material phase changes into a nematic phase, the smectic planes disappear but, in some places of the cell, these planes remain as little clusters, anchored at the cell walls. We have then these cybotactic clusters preserving a local smectic order. Moreover, defects on the walls can contribute to the persistence of these clusters. When the temperature further increases, the smectic order in cybotactic clusters is suppressed and the texture changes, more or less abruptly, in a smooth texture. The persistence of a local smectic order at the walls explains the memory effect observed in the cells of compounds C and D, if we suppose that the mechanism in the formation of the textures in these compounds is the same as in the alkyloxybenzoic materials.

The memory effect is also displayed in the cells prepared with compounds A and B, the materials with a more "smectic" behaviour. Samples E,F and G have only the nematic phase that possesses, on cooling, a wider range compared with that observed on heating. No memory effect is seen. The presence of cybotactic clusters can be excluded in E,F and G compounds. Or, it is better to tell in other words, cybotactic clusters do not grow with the decrease in the temperature, consistent with the fact that the medium does not support a smectic phase. This can be due to a kind of frustration developed in the medium, wherein a phase transition into layered structure is not possible due to the molecular structure. As it is shown in the molecular scheme of Fig.1, the responsible can be the position of the Cl atom.

All the results reported in this paper, on the smectic and nematic phases, confirm that the family of oxadiazoles is very promising for further researches in the investigation of texture transitions, and the relationship between the molecular structure and the features of the mesophases. Let us remember once more that two of these compounds shows a texture transition in the nematic range.

| Compound | Transition temperatures on heating |
|---|---|
| A | K-140°C-Sm-210°C-I |
| B | K-156°C-Sm-172°C-N-200°C-I |
| C | K-94°C-Sm-172°C-N-233°C-I |
| D | K-103°C-Sm-108°C-N-232°C-I |
| E | K-46°C-N-109°C-I |
| F | K-72°C-N-99°C-I |
| G | K-80°C-N-87°C-I |

**TABLE I: Transition temperatures of the compounds on heating. The first compound (A) has only the smectic mesophase. E,F and G possess only a nematic mesophase.**

**Figure captions**

Figure 1: Molecular structures of the seven oxadiazole compounds under investigation.

Figure 2: The crystal texture of compound A (upper part). Note the thick dark lines among the lamellar structure. These are cracks appearing on cooling from the smectic phase. In its lower part, the figure shows the bâtonnets growing in the isotropic melt on cooling (image dimensions 0.38mm x 0.5mm).

Figure 3: The sequence shows the cooling of compound A from the smectic fan-shaped texture in the lamellar texture of the crystal phase (image dimensions 0.38mm x 0.5 mm).

Figure 4: The sequence shows the cooling of compound B. In the upper part the nematic, in the middle the smectic and, in the lower part of the figure, the crystal phase.

Figure 5: The smectic phase of compound C. Note the toric focal conic domains. In the lower part of the figure, domains with undulations inside.

Figure 6: The transition from the smectic into the nematic phase of compound C: note the toric texture of the nematic phase.

Figure 7: The texture transition for sample C, on cooling. Note the growth of the low temperature nematic texture with the bent-shaped lines.

Figure 8: The toric nematic phase as it appears when heating the smectic phase of compound D.

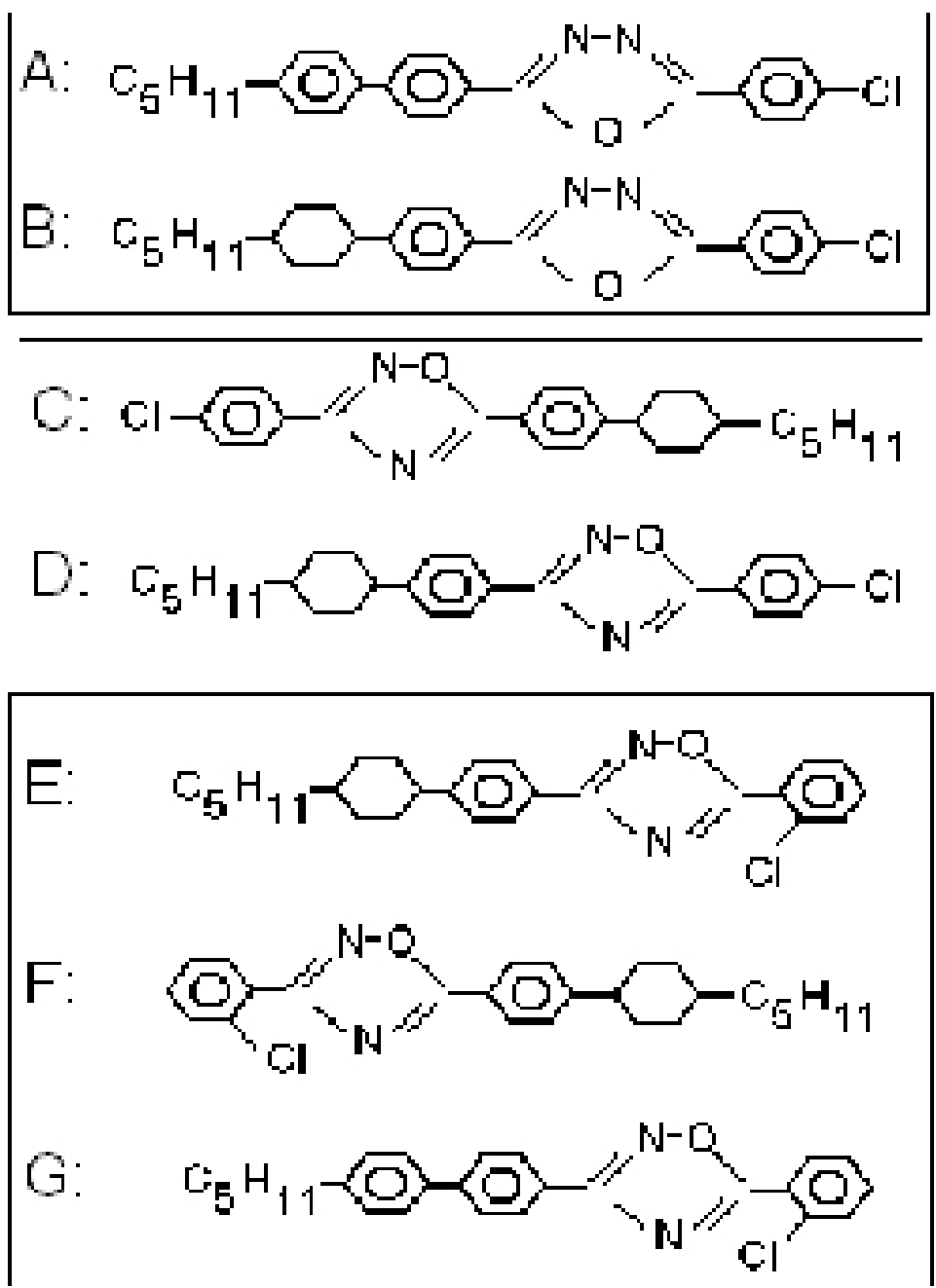

Figure 1

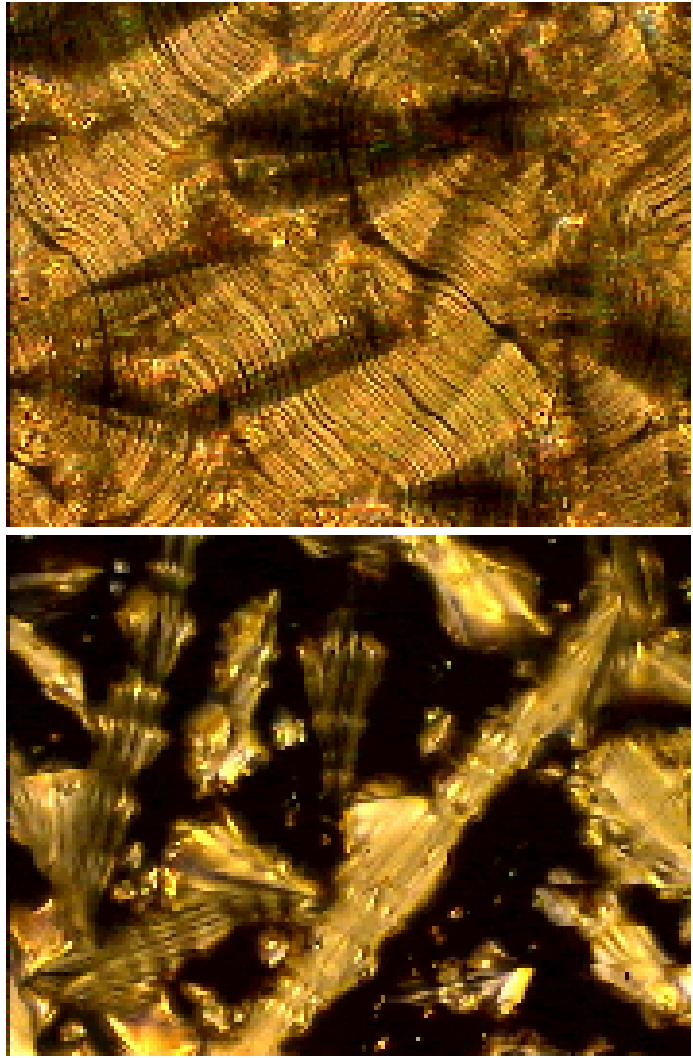

Figure 2

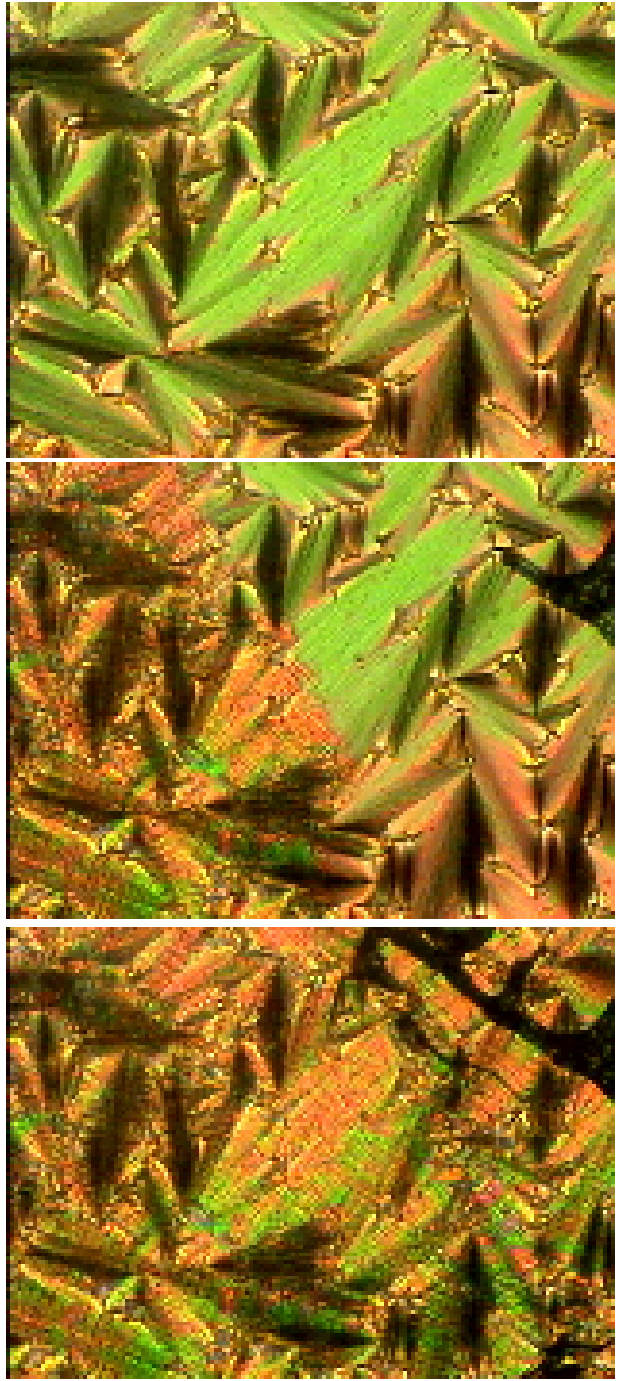

Figure 3

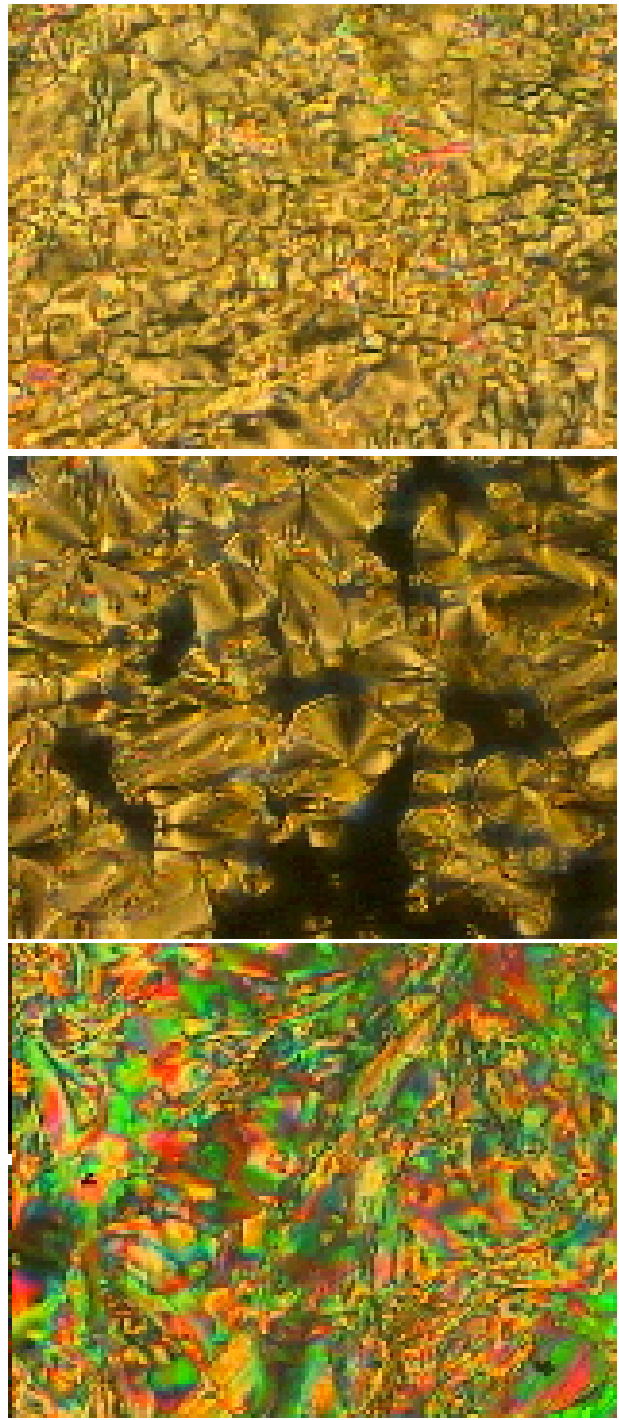

Figure 4

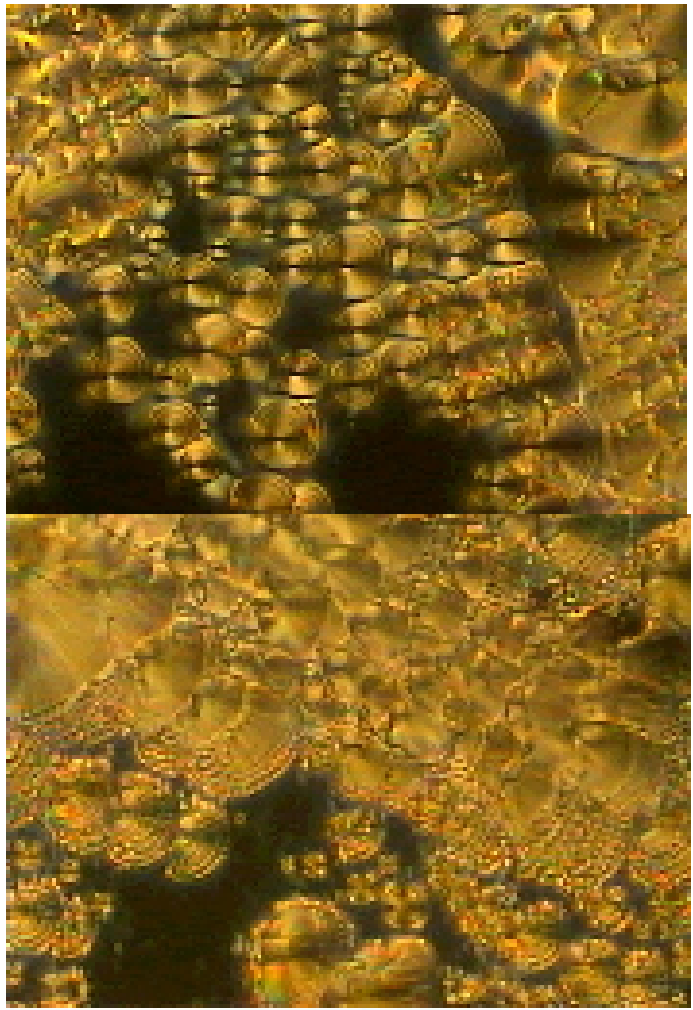

Figure 5

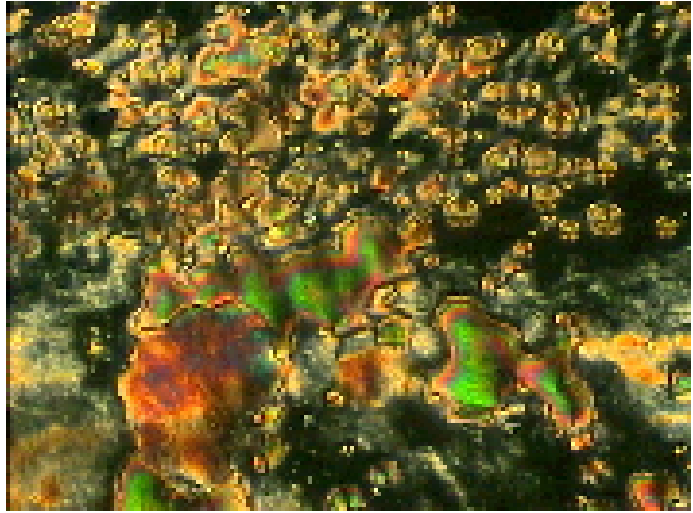

Figure 6

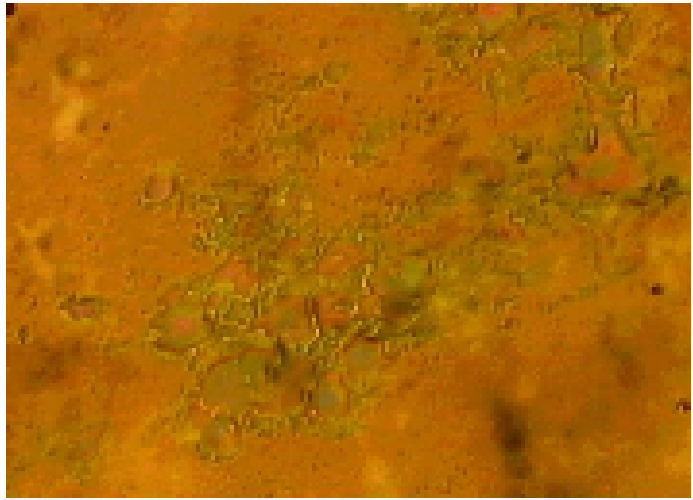
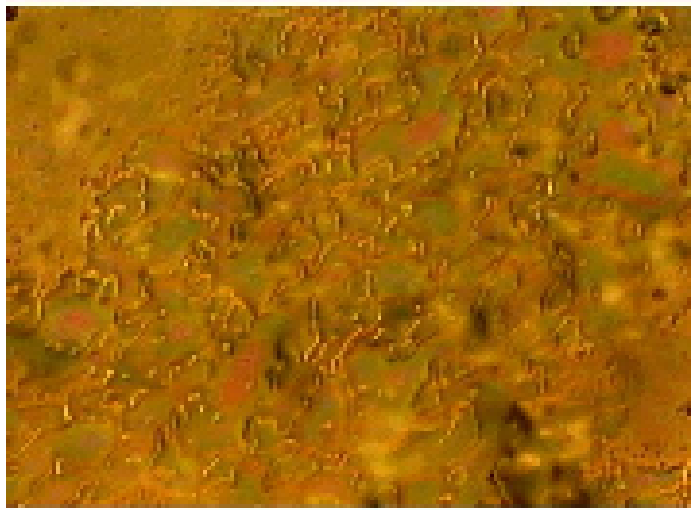

Figure 7

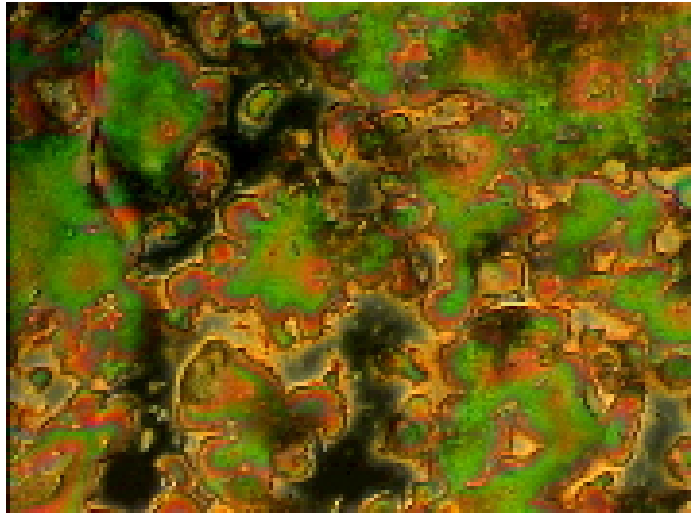

Figure 8